\documentclass[]{aa}  

\usepackage{natbib}
\bibpunct{(}{)}{;}{a}{}{,}
\usepackage{graphicx}
\usepackage{revsymb}	
\usepackage{amsmath}	
\usepackage{txfonts}
\usepackage{amssymb}
\usepackage{color}
\usepackage{adjustbox}
\usepackage{geometry} 
\usepackage{amssymb}
\usepackage{slantsc}
\usepackage{array}
\usepackage{url}
\usepackage{amsfonts}
\usepackage[colorlinks=true,linkcolor=black, urlcolor=black, citecolor=black]{hyperref}
\usepackage{hyperref} 
\hypersetup{citecolor=blue}
\hypersetup{linkcolor=blue}
\usepackage{graphicx}
\usepackage{xcolor}
\usepackage{nicefrac}
\usepackage{subfigure} 
\usepackage{epsfig}
\colorlet{rouge}{red!70!darkgray}
\usepackage{booktabs, caption} 

\begin{document}

\title{The key impact of the host star's rotational history on the evolution of \object{TOI-849b}}



\author{C. Pezzotti\inst{1} \and O. Attia\inst{1} \and P. Eggenberger\inst{1} \and G. Buldgen\inst{1}  \and {V. Bourrier}\inst{1}}
\institute{Observatoire de Genève, Université de Genève, Chemin Pegasi 51, CH$-$1290 Sauverny, Suisse\\
              \email{camilla.pezzotti@unige.ch}}


\date{Received ...; accepted ...}

 
 \abstract
   {TOI-849b is one of the few planets populating the hot-Neptune desert and it is the densest Neptune-sized one discovered so far. Its extraordinary proximity to the host star, together with the absence of a massive H/He envelope on top of the $40.8~M_{\oplus}$ rocky core, calls into question the role played by the host star in the evolution of the system.}
   {We aim to study the impact of the host star's rotational history on the evolution of TOI-849b, particularly focussing on the planetary migration due to dynamical tides dissipated in the stellar convective envelope, and on the high-energy stellar emission.}
   {Rotating stellar models of TOI-849 are coupled to our orbital evolution code to study the evolution of the planetary orbit. The evolution of the planetary atmosphere is studied by means of the JADE code, which uses realistic X-ray and extreme-ultraviolet (XUV) fluxes provided by our rotating stellar models.}
   {Assuming that the planet was at its present-day position ($a_{\rm in} = 0.01598$ AU) at the protoplanetary disc dispersal, with mass $40.8~M_{\oplus}$, and considering a broad range of host star initial surface rotation rates ($\rm \Omega_{\rm in} \in [3.2,18] ~\Omega_{\odot}$), we find that only for $\Omega_{\rm in} \leq 5~\Omega_{\odot}$ do we reproduce the current position of the planet, given that for $\Omega_{\rm in} > 5~\Omega_{\odot}$ its orbit is efficiently deflected by dynamical tides within the first $\rm \sim 40~$Myr of evolution. We also simulated the evolution of the orbit for values of $\rm a_{\rm in} \neq 0.01598$ AU for each of the considered rotational histories, confirming that the only combination suited to reproduce the current position of the planet is given by $a_{\rm in} = 0.01598~$AU and $\rm \Omega_{\rm in} \leq 5~\Omega_{\odot}$. We tested the impact of increasing the initial mass of the planet on the efficiency of tides, finding that a higher initial mass ($ M_{\rm in } = 1~M_{\rm Jup}$) does not change the results reported above. Based on these results we computed the evolution of the planetary atmospheres with the JADE code for a large range of initial masses above a core mass of $40.8~M_{\oplus}$, finding that the strong XUV-flux received by the planet is able to remove the entirety of the envelope within the first 50 Myr, even if it formed as a Jupiter-mass
planet.}
   {}

\keywords{Planet-star interaction - Planetary systems - Planets and satellites: atmospheres - Stars: evolution - Stars: rotation - Stars: solar-type}

\titlerunning{}
\maketitle

\section{Introduction}

\citet{Armstrong2020} recently announced the discovery of TOI-849b, a planet having a size comparable to the one of Neptune ($ R_{\rm pl} = 3.45^{+0.16}_{-0.12}~R_{\rm \oplus}$), but an anomalously larger mass ($ M_{\rm pl} = 40.8^{+2.4}_{-2.5}~M_{\oplus}$) and a density similar to the one of the Earth ($ \rho = 5.5 \pm 0.8~ \rm g/cm^{3}$). This is the densest Neptune-sized planet discovered to date. Interior structure models suggest that for this kind of planet, any H/He envelope would consist of no more than $\rm 3.9\%$ of the total mass \citep{Armstrong2020}. TOI-849b orbits around a late G-type star, with an orbital period of $ P = 18.4$ hours. Its equilibrium temperature is $ T_{\rm eq} = 1800$ K (for a Bond Albedo 0.3). With such properties, TOI-849b represents one of the few planets populating the hot Neptune desert, a region on the radius-orbital distance plane characterised by a surprising deficit of Neptune-sized planets on very short orbits \citep[e.g.][]{LecavelierdesEtangs2007, Beauge2013, Mazeh2016}. Growing evidence suggests that the evaporation of hot Neptunes due to stellar irradiation represents a major process in shaping the desert \citep{Owen2019}. Some of the planets within the desert or at its lower-radius border could thus be the remnant cores of larger gas-rich progenitors that lost most of their atmosphere \citep[e.g.][]{LecavelierdesEtangs2004}. Alternatively, orbital migration may have also played a part in sculpting the desert, with different classes of planets forming differently \citep[e.g.][]{Batygin2016}, or following different dynamical tracks \citep[e.g.][]{Matsakos2016}. Even so, the origin of this key feature in the demographics of close-in planets remains debated \citep{Zahnle2017, Owen2018}, and investigating the past history of planets such as TOI-849b contributes to the global effort towards disentangling the important mechanisms at the root of the desert. Since the mass of the planet is larger than the threshold value for runaway gas accretion \citep[roughly $ 10-20~M_{\oplus}$,][]{Mizuno1978,Rafikov2006,Piso2015}, TOI-849b might have been a gas giant before undergoing extreme mass loss via thermal self-disruption, collisions with other giant planets, or it was prevented from accreting gas because of gap openings in the protoplanetary disc, or because of late formation \citep{Armstrong2020}. In their work, \citet{Armstrong2020} found that their estimated photoevaporation rates cannot provide the mass-loss rates needed to remove a roughly $ 280~M_{\oplus}$ envelope from a Jupiter-like gas giant. Nevertheless, in their estimations the planetary atmosphere is not self-consistently monitored, and neither is the luminosity emitted by the host star relative to its rotational history. This is critical, considering how recent works emphasise the interplay between all these elements in driving the evolution of close-in worlds \citep[e.g.][]{Owen2016, Kubyshkina2018, King2021,Pezzotti2021}.

In this work we aim to study the orbital evolution of the system, investigating the impact of the host star's rotational history on the exchange of angular momentum between the star and the orbit through the dissipation of tides, and on the X-ray and extreme-ultraviolet (XUV) luminosities predicted for the host star. In this context, we also aim to follow the evolution of the planetary atmosphere and to provide estimations of the photo-evaporation rates using the XUV-luminosity evolutionary tracks derived from our rotating stellar models.

\section{Stellar model}

We computed rotating stellar models of TOI-849 by means of the Geneva stellar evolution code (GENEC) \citep{Eggenberger2008}. We used the values for the effective temperature ($ T_{\rm eff} = 5329 \pm 48 ~\rm K$), the iron abundance ($ \rm [Fe/H] = 0.2 \pm 0.03~ dex$), and the surface gravity ($\rm log~g = 4.43 \pm 0.3$) provided in \citet{Armstrong2020} as inputs and constraints to the models. Using the parallax values given by \textit{Gaia} \citep{Gaia2018}, we computed the bolometric luminosity ($ L/L_{\odot} = 0.568 \pm 0.027$) as in \citet{Buldgen2019}, applying the bolometric correction derived by the code of \citet{Casagrande2014,Casagrande2018a,Casagrande2018b} and the extinction inferred by the \citet{Green2018} dust map. We therefore derived a stellar mass of $ 0.93~M_{\odot}$, and a stellar radius of $ 0.91~R_{\odot}$ at an age of 6.45 Gyr, finding that these values are in good agreement with the ones in \citet{Armstrong2020}.

While a thorough description of the physics included in GENEC is provided in \citet{Eggenberger2008}, in the following we briefly recall the fundamental properties that are of interest for the present study. Rotational effects are accounted for in the hypothesis of shellular rotation \citep{Zahn1992}. The evolution of the star is followed simultaneously with the transport of angular momentum (AM), accounting for meridional currents, and transport by shear instabilities and magnetic fields. The transport by magnetic fields is taken into account in the context of the Tayler-Spruit dynamo \citep{Spruit2002}. In addition to the internal AM transport, we also account for the braking of the stellar surface rotation due to magnetised winds \citep{Matt2015, Matt2019}. 

\begin{figure}
\includegraphics[width=\linewidth]{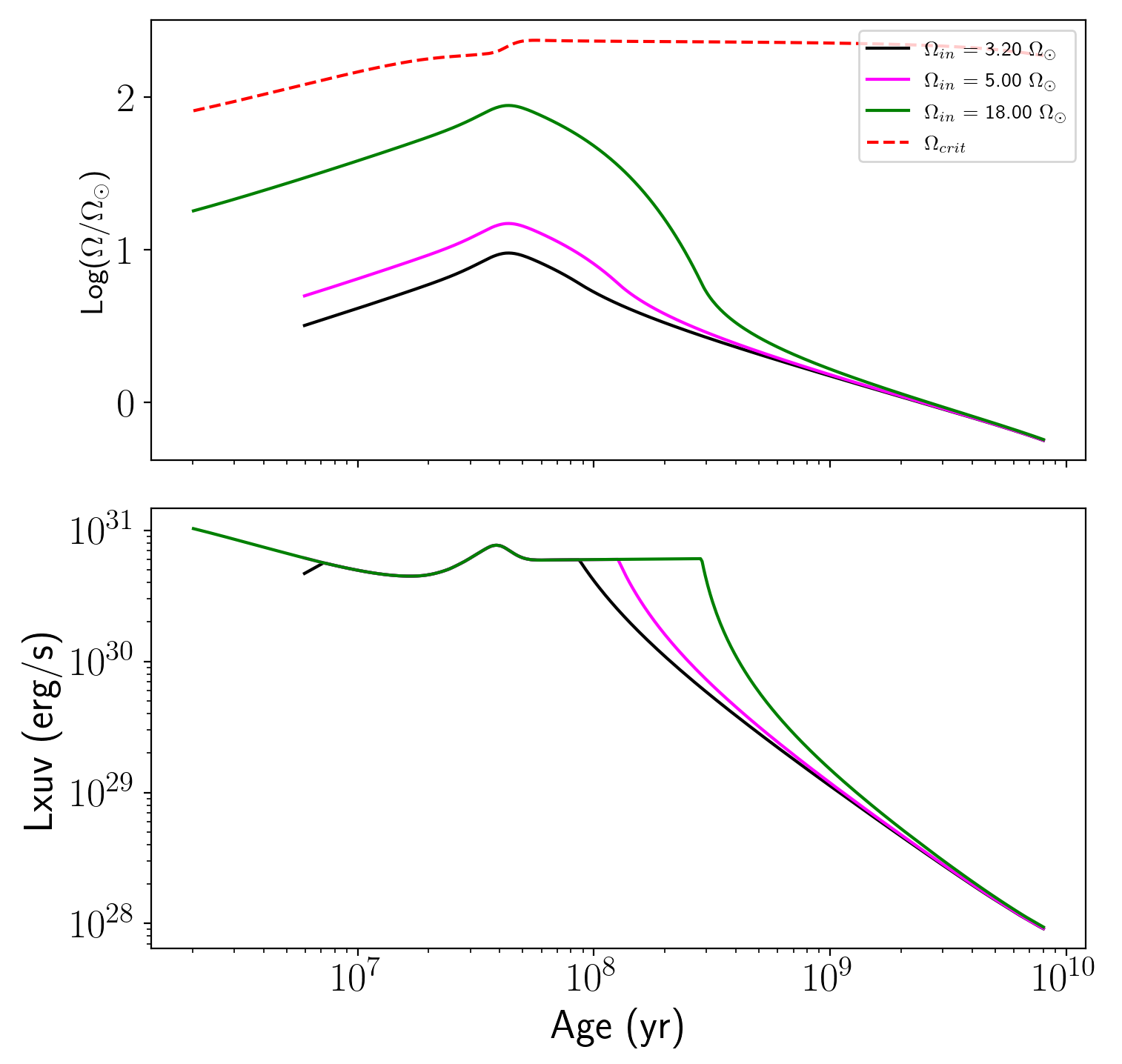}
\caption{\small{\emph{Top panel:} Evolution of the surface rotation rate of TOI-849 for a fast, medium, and slow rotator $\rm \Omega_{in} = 18, 5, 3.2~\Omega_{\odot}$ (green, magenta, and black line, respectively). The red-dashed line indicates the evolution of the critical rotation velocity ($\Omega_{\rm crit}$), defined as the velocity at which the centrifugal acceleration at the equator becomes equal to the  gravity.} \emph{Bottom panel:} Evolution of the XUV luminosities computed for each of the considered rotators. In both panels the starting point of the tracks corresponds to the dissipation of the protoplanetary disc (6 Myr for the slow and moderate rotators, 2 Myr for the fast one).}
\label{Omega_tot}
\end{figure}

With the rotational history of TOI-849 being unknown, we computed models representative of slow, medium, and fast rotators by using different values for the initial surface rotation rate ($\rm \Omega_{in} = 3.2, 5$ and $\rm 18~\Omega_{\odot}$, respectively) and for the disc lifetimes during the pre-main sequence (PMS) phase ($\rm \tau_{\rm dl} = 6$ Myr for the slow and medium rotators, $\rm \tau_{\rm dl} = 2$ Myr for the fast one) \footnote{The choice of these values for $\Omega_{\rm in}$ and $\tau_{\rm dl}$ was determined from the distribution of surface rotation rates observed for stars in star-forming regions and young open clusters at various ages \citep{Eggenberger2019a}.}. In the top panel of Fig.~\ref{Omega_tot}, we show the evolution of the surface rotation rate for each of the considered rotational histories. In the bottom panel of Fig.~\ref{Omega_tot}, the corresponding XUV-luminosity evolutionary tracks are represented. We computed the high-energy luminosity by recalibrating the prescription of \citet{Wright2011} for the X-ray and by using the one of \citet{SanzForcada2011} for the EUV luminosity \citep{Pezzotti2021}.

\section{Impact of the host star's rotational history on the evolution of the orbit}
\label{Sect:orbital_evolution}

In order to study the orbital evolution of TOI-849b, we coupled the rotating stellar models to our orbital evolution code \citep{Privitera2016AII,Privitera2016III,Rao2018}, and computed the total change of the orbital distance accounting for the dissipation of equilibrium and dynamical tides\footnote{Equilibrium tides represent the large-scale, non-wave-like flow induced in the host star by the presence of a planetary companion \citep{Zahn1966}; dynamical tides instead designate the wave-like solution of the gravitational response. In convective envelopes of rotating stars, these are constituted by inertial waves excited by the tidal potential, whose restoring force is the Coriolis force \citep{Zahn1975,Ogilvie2007}.} in the host star convective envelope along the evolution of the system. Thanks to this coupling, it is possible to study the AM exchange between the star and the planetary orbit and to test whether the dissipation of tides might have played a significant role in shaping the architecture of the system. Given the proximity of TOI-849b to the host star ($ a = 0.01598^{+0.00013}_{-0.00013}$ AU) and its relatively large mass ($ 40.8^{+2.4}_{-2.4}~ M_{\oplus}$), we may expect especially dynamical tides to have an impact in the early evolution of the system, with respect to a certain host star's rotational history

In our orbital evolution code, we consider the planet on a circular, coplanar orbit. The equation for the change of the orbital distance due to equilibrium tides is the same as in \citet{Privitera2016AII}. The contribution due to dynamical tides is included in the form of a frequency-averaged dissipation of inertial waves in the convective envelope of a rotating star and for their expression we refer to the works of \citet{Ogilvie2013} and \citet{Mathis2015}. The impact of tides is to widen (respectively shrink) the orbit of the planet when it is beyond (respectively inside) the corotation radius, defined as the distance at which the orbital period of the planet is equal to the stellar spin period, namely $a_{\rm corot} = [G(M_{\star} + M_{\rm pl})/\Omega_{\star}^2]^{1/3}$, where $G$ is the gravitational constant, $ M_{\star}$ is the mass of the star, $ M_{\rm pl}$ is the mass of the planet, and $\rm \Omega_{\star}$ is the host star surface rotation rate. In the context of the formalism used for dynamical tides, at each evolutionary time step, these are considered to be active only when the orbital distance is larger than a minimum critical value, defined as $ a_{\rm min} = 4^{-1/3}~a_{\rm corot}$ \citep{Ogilvie2007}. For a more detailed description of the tidal model used in the present work, we refer the interested reader to Appendix~\ref{App}. In the following we focus on describing the impact of dynamical tides on the planetary orbit since they represent the dominating mechanism governing the early evolution of the system when the star rotates much faster compared with more advanced evolutionary stages. 

\begin{figure}
\includegraphics[width=\linewidth]{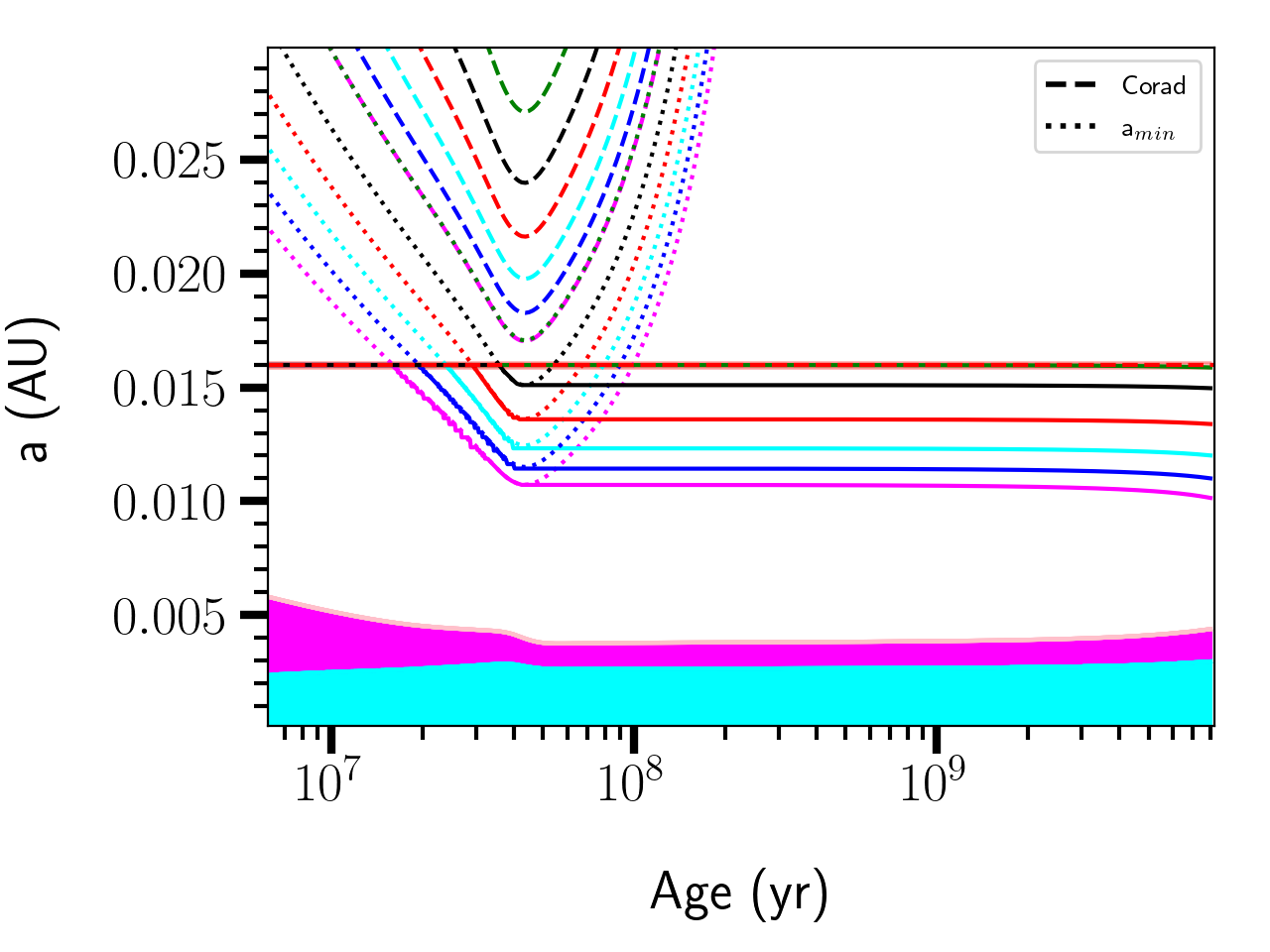}
\caption{\small{Evolution of the orbital distance for a planet having the same mass as TOI-849b and $ a_{\rm in} = 0.01598~ \rm AU$. From top to bottom, the solid lines show the evolution of the orbital distance for $\rm \Omega_{in} = 5, 6, 7, 8, 9, 10~\Omega_{\odot}$ (green, black, red, cyan, blue, and magenta). The dotted and dashed lines indicate the evolution of $ a_{\rm min}$ and $a_{\rm corot}$ (or Corad), respectively, for different values of $\rm \Omega_{in}$ (same colour code). The cyan and magenta shaded areas represent the extension of the stellar radiative and convective regions, respectively.}}
\label{max_Omega}
\end{figure}

As a first step, we computed the orbital evolution of the planet at a constant planetary mass ($ 40.8~M_{\oplus}$), assuming that it started the evolution with an initial orbital distance equal to its present-day value ($ a_{\rm in} = 0.01598$ AU). Since the rotational history of the star is unknown, we simulated the evolution of the orbit considering different values for the host star's initial surface rotation rate ($\Omega_{\rm in} = 3.2, 5, 18~\Omega_{\odot}$). In this way, we tested for the considered rotational histories whether the planet would remain at its initial (and present-day) orbital distance until an age of approximately 6.45 Gyr. We found that for $\Omega_{\rm in} = 3.2, 5~\Omega_{\odot}$, there is no significant impact on the evolution of the orbit; instead for $\Omega_{\rm in} = 18~\Omega_{\odot}$, the orbit of the planet rapidly shrinks on timescales of the order of $\rm \sim 40$ Myr. At this stage, given the proximity of the planet to the star, it might get totally disrupted by tides or get engulfed by the host star at $\sim 400$ Myr. Given the significant change in the dynamical tides efficiency occurring between the moderate and the fast rotator case, we considered a series of rotators with $\Omega_{\rm in} \in [5,18]~\Omega_{\odot}$, in steps of $1~\Omega_{\odot}$, to find the value of $\Omega_{\rm in}$ (and the corresponding rotator) below which dynamical tides leave the planetary orbit unperturbed. We found that for $\rm  \Omega_{in}  \leq 5~\Omega_{\odot}$, there is no significant impact on the evolution of the orbit; instead, for $\rm \Omega_{in} > 5~\Omega_{\odot}$, the orbit of the planet shrinks on timescales of the order of $\rm 40$ Myr, leading us to obtain final values that do not reproduce the current position of the planet. The results related to these simulations are shown in Fig.~\ref{max_Omega}. We recall that in the context of the formalism used in the present work, at each time step dynamical tides are considered to be active only for $a \geq a_{\rm min}$. Given the dependency of $ a_{\rm min}$ on $\rm \Omega_{\star}$, we notice that for values of $\rm \Omega_{in} \leq 5~\Omega_{\odot}$, if we start the simulation at $a_{\rm in} = 0.01598$ AU, the orbit of the planet remains below $ a_{\rm min}$ for the whole evolution of the system, and consequently dynamical tides never become active. Instead, for values of $\rm \Omega_{in} > 5~\Omega_{\odot}$, either the value of $a_{\rm min}$ is smaller than $\rm 0.01598$ AU from the beginning of the evolution (for $\rm \Omega_{in} \geq 16~\Omega_{\odot}$), or it becomes smaller at $\rm\sim 20-30$ Myr, allowing dynamical tides to be active. Therefore, regarding the hypothesis that TOI-849b started its evolution at $\rm 0.01598$ AU and evolved at constant mass, our results show that only for $\rm \Omega_{in} \leq 5~\Omega_{\odot}$ are we able to reproduce its current position.

\begin{figure}
\includegraphics[width=\linewidth]{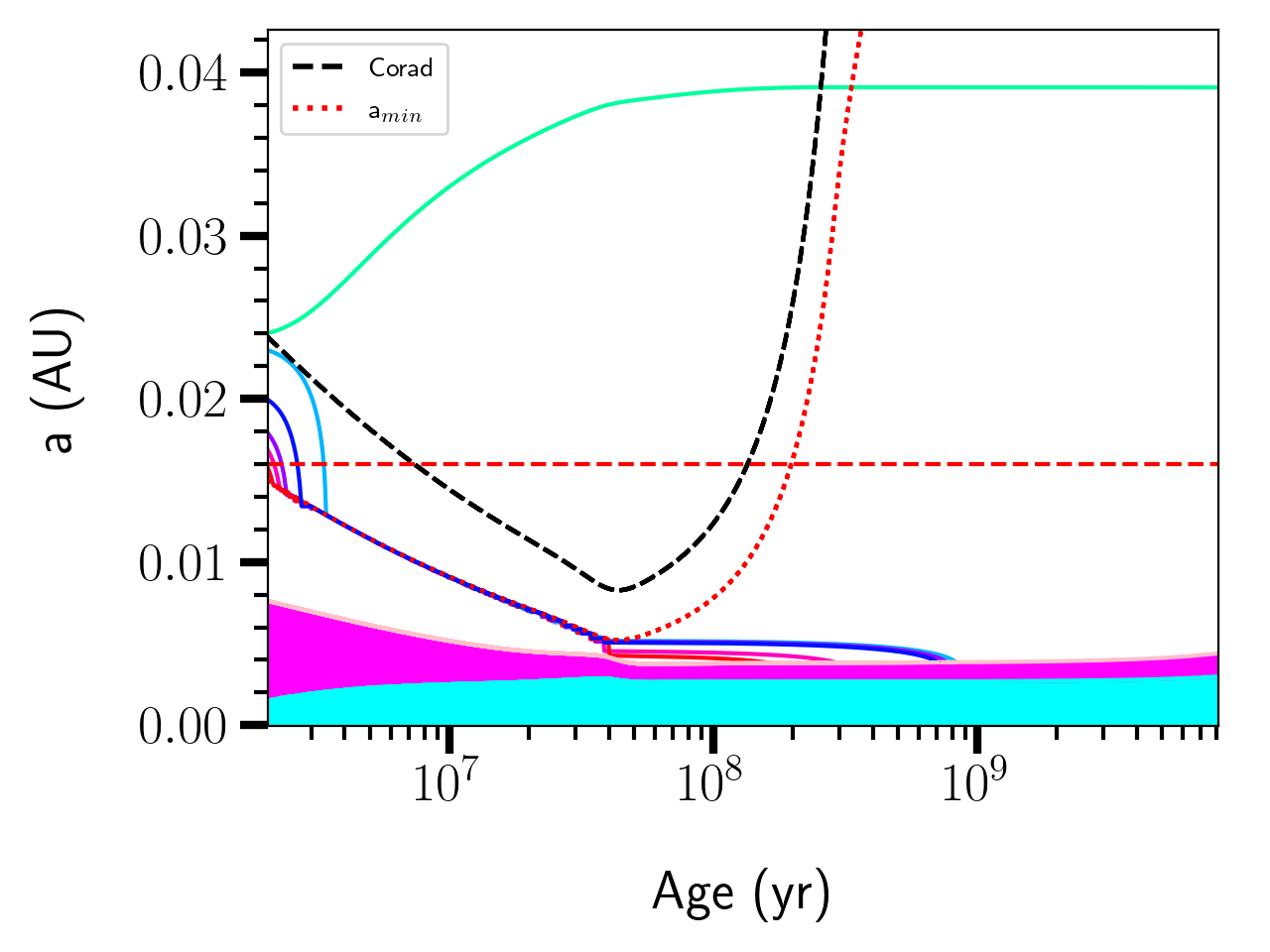}
\caption{\small{Evolution of the planetary orbital distance (solid lines) in the case of a fast rotator ($\rm \Omega_{in} = 18~\Omega_{\odot}$), spanning a range of initial values $\rm 0.01598 - 0.025~AU$. The red-dashed line represents the current orbital distance of TOI-849b. The red-dotted and black-dashed lines show the evolution of $ a_{\rm min}$ and $a_{\rm corot}$, respectively. The cyan- and magenta-shaded areas represent the extension of the stellar radiative and convective regions, respectively.}}
\label{18_Omega}
\end{figure}

We may ask whether for $\rm \Omega_{in} > 5~\Omega_{\odot}$ can the present position of the planet be reproduced, this time changing its initial orbital distance (namely for $ a_{\rm in} \neq 0.01598$ AU). Nevertheless, we found that there is no value of $ a_{\rm in}$ for which this is possible. This is shown in Fig.~\ref{18_Omega} for the specific case of the fast rotator ($\rm \Omega_{in} = 18~\Omega_{\odot}$), for which the impact of tides is the most evident. 

In the previous computations, we assumed that the planet evolved at a constant mass ($ 40.8~M_{\oplus}$). Nevertheless, TOI-849b might have formed with a significantly higher mass \citep[roughly $ 1~M_{\rm Jup}$, ][]{Armstrong2020}. Therefore, for each of the considered rotational histories, we recomputed the evolution with $M_{\rm in} = 1~M_{\rm Jup}$, and different values of $a_{\rm in}$, finding that assuming a higher initial mass does not change the results we reported in the previous paragraphs. This shows that for the assumed configuration of the system, the rotational history of the host star mainly determines the efficiency of tides.

The reproducibility of the planetary position at $\rm \sim 6.45~Gyr$ for values of $\rm \Omega_{in} \leq 5~\Omega_{\odot}$ helps to reduce the degeneracy on the host star's rotational history, indicating that an evolution as a slow or moderate rotator might be favoured over the fast one. This result also provides constraints on the high-energy emission history of the host star, with its activity and rotational history being directly linked. In this context, given that the planet likely remained its entire life at its present-day position if it migrated early on, and that its initial mass does not represent a dominating factor for the efficiency of tides, it is possible to  simulate the atmospheric evolution of the planet separately, considering  different initial masses at a fixed orbital distance.

\section{Impact of evaporation}

We study the impact of atmospheric processes on the evolution of the mass of the planet. To this effect, we employed the JADE code \citep{Attia21}. We refer to that paper for details, but briefly summarise the
main ingredients here. The JADE code allows for one to simulate the evolution of a planet over secular timescales under the coupled influence of complex dynamical and atmospheric mechanisms. The planet is composed of a rocky core topped by a H/He gaseous envelope. The planet's radius is self-consistently derived by integrating its thermodynamical structure as it varies following the evolution of the orbit, the amount of stellar irradiation, and inner planetary heating. The atmosphere is prone to erode due to XUV-induced photo-evaporation, with a mass-loss rate calculated as \citep[e.g.][]{Watson1981,Lammer2003,Jin2014}

\begin{equation}
    \dot{M}_\mathrm{env} = \epsilon \left\langle F_\mathrm{XUV} \right\rangle \frac{S_\mathrm{XUV}}{\Phi_0 K_\mathrm{tide}},
\end{equation}
where $M_{\rm env}\equiv{}M_{\rm pl}-M_{\rm core}$ is the mass of the gaseous envelope, $\epsilon$ the efficiency of photo-evaporation, $F_{\rm XUV}$ the stellar XUV flux, $S_{\rm XUV}$ the cross-section surface that collects the high-energy irradiation, $\Phi_0$ the planetary gravitational potential, and $K_{\rm tide}$ a correction factor accounting for atmospheric loss enhancement due to the action of tidal forces \citep{Erkaev07}. One of the novelties of the JADE code is that the XUV flux is consistently secularised  with the dynamical equations of motion, so as to better account for the orbital evolution \citep{Attia21}. The factors $\epsilon$ and $R_{\rm XUV}$ are computed using analytical formulae derived from self-consistent simulations of upper atmospheric structures \citep{Salz16}, which allows a better agreement with observed mass-loss rates than with the simple energy-limited approach.

To investigate the sole impact of photo-evaporation in accordance with the results found in the previous section, we conducted purely atmospheric simulations at a fixed circular orbit ($a_{\rm in}=0.01598~{\rm AU}$), for two sets of simulations corresponding to the star being a slow or an intermediate rotator ($\Omega_{\rm in}=3.2, 5 \times\Omega_\odot$). 
We used an atmospheric helium fraction $Y=0.15$ as a typical value for Neptune-sized planets \citep[e.g.][]{Hubbard95}. Since \citet{Armstrong2020} constrained a maximum H/He envelope mass fraction of 3.9\%, we considered a rocky core mass of $M_{\rm core}=40.8~M_\oplus$, which corresponds to the currently measured value of the total mass. We simulated the atmospheric evolution of the planet for a large range of initial masses above this value.

\begin{figure}
\centering
\includegraphics[width=8.5cm]{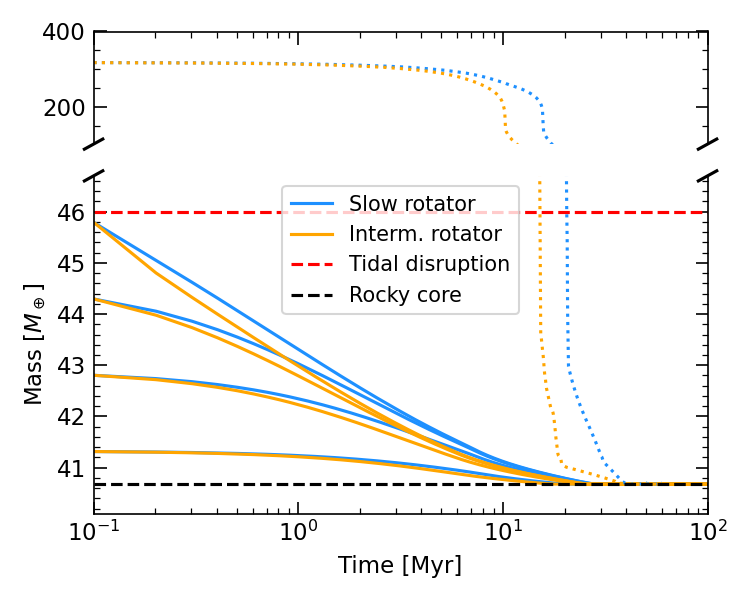}
\caption{\small{Evolution of the planetary mass under the influence of photo-evaporation for different initial mass values. Blue (respectively orange) curves correspond to the star being a slow (respectively intermediate) rotator. The red-dashed line indicates the mass threshold above which the planet would be immediately tidally disrupted. The black-dashed line indicates the mass of the rocky core. The dotted lines correspond to simulations of a $1~M_{\rm Jup}$ initial mass planet without accounting for the tidal disruption limit. We point out the broken $y$-axis. The starting time of the evolution ($t=0$) corresponds to the protoplanetary disc dispersal.}}
\label{fig:massloss}
\end{figure}

Figure \ref{fig:massloss} illustrates the results. Simulated planets with an initial mass higher than $\sim 46~M_\oplus$ are affected by tidal disruption. Indeed, for these large atmospheric masses and at such short ranges from the star, the planetary radius is highly inflated ($R_{\rm pl} \gtrsim 1~R_{\rm Jup}$). The atmosphere then extends behind the Roche limit of the star \citep[e.g.][]{Gu03,Jackson17} and we assume it is completely depleted by disruption due to tidal interactions. For initial masses below this threshold, the planet always loses the entirety of its atmosphere within $\sim 30~{\rm Myr}$ from the dissipation of the protoplanetary disc due to photo-evaporation leaving a rocky bare core. This result is valid for both rotators, with a slightly faster evaporation for the intermediate rotator as in this case the high-energy flux received by the planet is higher (see Fig.~\ref{Omega_tot}).

Tidal atmospheric disruption is likely more complex than the simple threshold condition used in our simulations. To investigate whether a more massive planet could keep its atmosphere if it was more resilient to disruption, we considered the extreme case where tidal disruption is inactive and the planet only evolves under photo-evaporation. We simulated the case where TOI-849b formed as a Jupiter-mass planet, unaffected by disruption. The result, illustrated as dotted lines in Fig.~\ref{fig:massloss}, shows that the gaseous envelope still completely erodes after $\sim 40$ Myr. Consequently, the efficient removal of the atmosphere, even for a high initial planetary mass, is a robust result regardless of whether the Roche limit is considered or not.

Given the very close orbit of TOI-849b and high stellar emission of its host star after formation, the XUV flux received by the planet is one order of magnitude higher than the range of validity of our mass loss formalism from \citet{Salz16}. However, given the speed of TOI-849b's full atmospheric erosion, a possible overestimation of the mass loss rate does not change our conclusion that the present-day planet would have lost its atmosphere long ago if it had evaporated following its formation. Several scenarios can be put forward to explain its possible residual gaseous envelope. A simple solution could be a different planetary composition \citep[e.g. water-dominated,][]{Lopez17, Armstrong2020}, which would allow for a wider range of initial masses with compact atmospheres more resilient to tidal disruption and photo-evaporation. Alternatively, a different atmospheric history can be envisioned in the framework of late orbital migration. If the planet formed far enough from its star and did not migrate early on, it would have been largely spared from photo-evaporation during the first $\sim 100~{\rm Myr}$, when stellar emission is the highest (see Fig.~\ref{Omega_tot}). Migrating billions of years later to its current orbit, the planet would have then been subjected to a stellar irradiation decreased by several orders of magnitude. We will investigate this scenario, which could explain how TOI-849b still holds a volatile atmosphere, in a follow-up paper.

\section{Conclusion}

In this work we aimed to study the impact of the stellar rotational history on the past evolution of the TOI-849b, considering both the impact of tides on the planetary orbit and the irradiation of the planetary atmosphere. Coupling the rotating stellar models computed with the Geneva code to our orbital evolution code, assuming that the planet was at $ 0.01598$ AU and had its present-day mass ($40.8 ~M_{\oplus}$) at the dispersal of the protoplanetary disc, we found that the impact of dynamical tides on the evolution of the orbit becomes significant only for values of the surface rotation rate of the star ($\Omega_{\rm in}$) above $\rm 5~\Omega_{\odot}$. For these values of $\Omega_{\rm in}$, the planet migrates towards the host star during the first $1 - 100$ Myr and it is eventually engulfed later on. Relaxing the assumption of an initial orbital distance at $0.01598$ AU, we explored a range of values for $a_{\rm in}$ and $\Omega_{\rm in}$ in order to find a combination for which it is possible to reproduce the current position of the planet thanks to the impact of tides. We found that such a configuration is only retrieved for $a_{\rm in} = 0.01598$ AU and $\Omega_{\rm in} \leq 5~\Omega_{\odot}$. Simulations performed for a higher initial planetary mass, up to $1~M_{\rm Jup}$, and various values of $a_{\rm in}$ and $\Omega_{\rm in}$ do not change this conclusion.


Thanks to the constraints derived on the host star's rotational history, we selected a set of XUV-luminosity evolutionary tracks computed consistently with the evolution of the surface rotation of the star to study the impact of photo-evaporation on the planetary atmosphere. We performed atmospheric simulations with the JADE code assuming a fixed circular orbit at $ 0.01598$ AU, a H/He envelope, and an in situ formation or early-on migration, and we computed the evolution of the planetary atmosphere for a large range of initial masses above a core mass of $40.8~M_{\oplus} $. The strong irradiation associated with the saturated regime of emission of the host star leads TOI-849b to completely evaporate its atmosphere in less than 50 Myr, even if it had formed as a Jupiter-mass planet. In addition, initial planetary masses above $46~M_\oplus$ are associated with highly inflated radii, so much so that the atmosphere would be affected by tidal disruption as it extends beyond the stellar Roche limit. We conclude from these atmospheric simulations that TOI-849b could not have retained a fraction of its gaseous envelope if it reached its current orbit just after the dissipation of the protoplanetary disc. If TOI-849b still has a residual atmosphere today, it likely has a peculiar atmospheric composition, or it underwent a delayed migration. In the latter scenario, the planet would have avoided erosion by remaining far from the star during the first $\sim 100$ Myr, when the stellar emission is the strongest. 

This study demonstrates how thoroughly considering the rotational history of a host star together with the evolution of the planet may lead to alternative results that would be unexplored otherwise. We indeed found that in the configuration assumed for this system, it is possible for a planet such as TOI-849b to lose the entirety of its envelope even if it formed as a Jupiter-mass planet. A further detailed investigation of this system, which will be carried out in a follow-up paper, will eventually help to shed light on the relevant mechanisms shaping the planets populating the hot Neptune desert and determine their possible filiation.

\section*{Acknowledgements}
We thank the referee for the useful comments that helped to improve the quality of the manuscript. G.B. acknowledges funding from the SNF AMBIZIONE grant No 185805 (Seismic inversions and modelling of transport processes in stars). P.E. has received funding from the European Research Council (ERC) under the European Union's Horizon 2020 research and innovation program (grant agreement No 833925, project STAREX). This project has received funding from the ERC under the European Union’s Horizon 2020 research and innovation program (project SPICE DUNE, grant agreement No 947634). V.B. \& O.A. acknowledge support by the Swiss National Science Foundation (SNSF) in the frame of the National Centre for Competence in Research “PlanetS”.

\bibliographystyle{aa}
\bibliography{biblioarticleTOI}

\begin{appendix}
\section{Equilibrium and dynamical tides model}
\label{App}

In our orbital evolution code \citep{Privitera2016AII,Privitera2016III,Rao2018}, we computed the total change of the planetary orbital distance by accounting for the dissipation of equilibrium and dynamical tides in the host star. While equilibrium tides constitute the non-wave-like large-scale flow induced in the star by the presence of the planet, dynamical tides instead represent the wave-like solution of the gravitational response. 
We coupled rotating stellar models that were computed with the Geneva stellar evolution code to the orbital evolution code in order to take into account the evolution of the fundamental stellar parameters and their impact on the tidal dissipation efficiency. 

Equilibrium tides are mainly dissipated in the low-mass star convective envelopes through turbulent friction induced by convection \citep{Zahn1966}, while their damping is less efficient in other regions \citep{Zahn1977}. In our approach, we therefore account for the dissipation of equilibrium tides only when a convective envelope is present. The equation determining the change of the orbital distance due to equilibrium tides reads as \citep{Privitera2016AII}

\begin{equation}
 \left( \dot{a}/a \right)_{\rm eq} = \frac{f}{\tau} \frac{M_{\rm env}}{M_{\star}} q(1+q) \left( \frac{R_{\star}}{a} \right)^8 \left( \frac{\Omega_{\star}}{\omega_{\rm pl}} - 1\right),
\end{equation}
\label{eq:equi_tides}

where the term $a$ is the orbital distance and $\dot{a}$ is its derivative with respect to time. The term $ f$ is a numerical factor obtained from integrating the viscous dissipation of the tidal energy across the convective zone \citep{Villaver2009}, which is equal to 1 when the convective turnover timescale is larger than half of the orbital period ($\tau > P_{\rm orb}/2$), otherwise  $f = (P_{\rm orb}/ 2\tau)^2$ \citep{Goldreich1977}. The factor $M_{\rm env}$ is the mass of the convective envelope, $ q$ is the ratio between the mass of the planet and the one of the star ($ q = M_{\rm pl}/M_{\star}$), $ \Omega_{\star}$ is the stellar surface rotation rate, and $ \omega_{\rm pl} = 2\pi/ P_{\rm orb}$ is the orbital frequency of the planet. 

Dynamical tides might be efficiently dissipated both in convective and radiative regions of a star. In convective regions, they are constituted by inertial waves driven by the Coriolis force. These waves are excited when the values of the tidal frequency ($\omega_t$) are comprised in the range $ [-2 \Omega_{\star}, 2 \Omega_{\star}]$  \citep{Ogilvie2007}. In our code the impact of dynamical tides propagating in convective zones is accounted for in the form of a frequency-averaged tidal dissipation of inertial waves \citep{Ogilvie2013,Mathis2015,BolmontMathis2016}. This expression is considered assuming a two-layer (core and envelope) model for the host star, each of them characterised by a uniform density \citep{Mathis2015}. We note that assuming this schematic stratification for the stellar structure, together with the use of a frequency-average dissipation rate for the dynamical tide, adds a certain degree of uncertainty on our results. Nevertheless, while this approach suffers from some schematic simplifications, on the other hand it has the advantage of providing us with relevant orders of magnitude of tidal dissipation rates accounting for the evolution of the structural and rotational parameters of the host star \citep{Mathis2015, BolmontMathis2016, Rao2018, Barker2020}. 
If we assume that the planet is on a circular-coplanar orbit, dynamical tides are considered to be active whenever the condition $\rm \omega_{pl} < 2~\Omega_{\star}$ is satisfied, since in this configuration the planet is able to excite inertial waves in the stellar convective envelope only when the orbital frequency is lower than twice the stellar rotation rate \citep{Ogilvie2007}. The equation defining the change of the orbital distance due to the dissipation of dynamical tides \citep{Ogilvie2013, Mathis2015} reads as

\begin{equation}
\left( \dot{a}/a \right)_{\rm dyn} = \left( \dfrac{9}{2Q^{\prime}_{\rm d}} \right)q \omega_{\rm pl} \left( \frac{R_{\star}}{a} \right)^5 \dfrac{(\Omega_{\star} - \omega_{\rm pl})}{\mid \Omega_{\star} - \omega_{\rm pl}\mid},
\end{equation}
\label{eq:dyn_tides}

where $ Q^{\prime}_{\rm d} = 3/(2D_{\omega})$ and  $\rm D_{\omega} = D_{0\omega}D_{1\omega}D_{2\omega}^{-2}$. The `D' terms are defined as

\begin{equation}
\begin{cases}
D_{0\omega} = \dfrac{100\pi}{63} \epsilon^{2} \dfrac{\alpha^5}{1 - \alpha^5} (1 - \gamma)^2,\\
D_{1\omega} = (1 - \alpha)^4 \left( 1 + 2\alpha + 3\alpha^2 + \frac{3}{2} \alpha^3 \right)^2 ,\\
D_{2\omega} = 1 + \frac{3}{2}\gamma + \frac{5}{2 \gamma}\left( 1 + \frac{\gamma}{2} - \frac{3 \gamma^2}{2} \right) \alpha^3 - \frac{9}{4}\left(1 - \gamma\right)\alpha^5  ,
\end{cases}
\end{equation}

where $\alpha = R_{\rm c}/R_{\star}$, $\beta = M_{\rm c}/M_{\star}$, $\gamma = \dfrac{\alpha^3 (1 - \beta)}{\beta (1 - \alpha^3)}$,  and$\epsilon = \dfrac{\Omega_{\star}}{\sqrt{\dfrac{GM_{\star}}{R_{\star}^3}}}$. The terms $M_{c}$ and $R_{c}$ indicate the mass and radius of the radiative core. The term $\rm D_{\omega}$ was computed in \citet{Ogilvie2013} as the frequency-averaged tidal dissipation. 

It is possible that under specific conditions, waves are excited at the convective-radiative interface and that they propagate in the radiative regions as internal-gravity waves, eventually contributing to the dissipation of energy and the exchange of angular momentum between the planet and the host star \citep[e.g.][]{Goodman1998,Ogilvie2007,Barker2010, Barker2020}. In the current version of our orbital evolution code, we do not account for the dissipation of dynamical tides in stellar radiative regions. In order to understand the implications of not considering this dissipation mechanism in the orbital history of TOI-849b, we analysed the results recently obtained in the work of \citet{Barker2020}. In this work, the dissipation of dynamical tides in radiation zones is computed accounting for the evolution of the relevant stellar quantities. \citet{Barker2020} assumes that the internal gravity waves are launched from the convective-radiative interface, and that these are subsequently fully damped. Such an efficient damping is likely justified when wave breaking occurs; also, in order for this mechanism to be triggered, it is necessary that the planetary mass is above a certain critical value ($M_{\rm critic}$). \citet{Barker2020} studied the evolution of $M_{\rm critic}$ with respect to the evolution of solar-like and low-mass stars, finding that for stars with masses lower than $1~M_{\odot}$, the planetary critical mass goes below $1~M_{\rm Jup}$ only at ages larger than about 10 Gyr. This result indicates that only planets more massive than $1~M_{\rm Jup}$ are able to trigger wave breaking during the PMS and most of the MS phases for these stars. Nevertheless, as mentioned before, the value of $M_{\rm critic}$ rapidly drops below $1~M_{\rm Jup}$ when these stars are close to the end of the MS, and subsequently super-Earths and mini-Neptunes may be massive enough to trigger wave breaking once their host stars are in the sub-giant and red giant branch phases. 

We recall that in the case of the TOI-849 system, the host star is a main sequence star with mass $0.93~M_{\odot}$ and $ \rm [Fe/H] = 0.2 \pm 0.03~ dex$. In order to get an estimation of the order of magnitude required for $M_{\rm critic}$ to trigger wave breaking, and recalling that the metallicity of TOI-849 is super solar, we examined the tracks representing the evolution of $M_{\rm critic}$ as a function of the stellar mass and age reported in Fig. 9 of \citet{Barker2020}. Indicatively, we focussed on the region comprised between the tracks representing the evolution of $M_{\rm critic}$ with respect to the $1.0$ and $0.8~M_{\odot}$ stellar models. We noticed that $M_{\rm critic}$ reaches values larger that about $1000~M_{\rm Jup}$ at the beginning of the evolution and it remains largely above $1~M_{\rm Jup}$ for the whole PMS and most of the MS phase, while it drops below $1~M_{\rm Jup}$ at approximately $7$ and $20$ Gyr for the $1$ and $0.8~M_{\odot}$ tracks, respectively. If we assume that wave breaking is the only mechanism allowing for a full dissipation of progressive internal gravity waves, in the case of TOI-849 it seems unlikely that such a process occurred during the early stages of evolution because it would have required TOI-849b to have been formed with a mass of the order of $1000~M_{\rm Jup}$. It is also unlikely that wave breaking occurred during the first part of the MS experienced so far by the star, because even if TOI-849b formed with a mass of about $1-2~M_{\rm Jup}$, this would have not been sufficient to trigger this process. On the contrary, it is possible that this process will be triggered at more advanced evolutionary stages. 

According to these considerations, it seems unlikely that the dissipation of dynamical tides in the radiative core of TOI-849 might have influenced the past history of the system. The definition of a model able to determine the secular evolution of the planetary system, accounting for the dissipations of tides in stellar radiative and convective zones in a consistent way with respect to the evolution of the host star, will allow us to explore in more detail the dependencies of planetary migrations with respect to these dissipations processes in the future.

\end{appendix}

\end{document}